\input{aipcheck}

\documentclass[
    ,final            
  ]
  {aipproc}

\layoutstyle{6x9}

\newcommand{\bce}{\begin{center}} 
\newcommand{\ece}{\end{center}}
\newcommand{\beq}{\begin{equation}}
\newcommand{\eeq}{\end{equation}}
\newcommand{\bea}{\vspace{0.25cm}\begin{eqnarray}}
\newcommand{\eea}{\end{eqnarray}}

\newcommand{\do}{\!\downarrow}

\newcommand{\ba}{\begin{array}}
\newcommand{\ea}{\end{array}}

\def\lsim{\mathrel{\rlap{\lower4pt\hbox{\hskip1pt$\sim$}}
    \raise1pt\hbox{$<$}}}         
\def\gsim{\mathrel{\rlap{\lower4pt\hbox{\hskip1pt$\sim$}}
    \raise1pt\hbox{$>$}}}         

\def\Pom{{\bf I\!P}}

\def\beq{\begin{equation}}
\def\endeq{\end{equation}}
\def\arr{\begin{eqnarray}}
\def\endarr{\end{eqnarray}}

\begin{document}

\title{BFKL, BK and the Infrared}

\classification{13.15.+g 13.60.Hb}
\keywords      {small-x evolution, non-linear effects,
infrared regularization}

\author{R.~Fiore}{
  address={Dipartimento di Fisica,
Universit\`a     della Calabria
and
INFN, Gruppo collegato di Cosenza,
I-87036 Rende, Cosenza, Italy}
}

\author{ P.V.~Sasorov}{
  address={ITEP, Moscow 117218, Russia}
}

\author{ V.R.~Zoller}{
  address={ITEP, Moscow 117218, Russia}
}

\begin{abstract}
The  perturbative non-linear (NL) effects in the small-$x$
  evolution of the  gluon densities depend crucially on the infrared
 (IR) regularization. The IR regulator, $R_c$, is determined by 
the scale of the non-perturbative fluctuations of QCD vacuum. 
From the instanton models and from the lattice $R_c\sim 0.3$ fm.
For perturbative gluons with the propagation length $R_c= 0.26$ fm
the linear BFKL  gives a good description of the 
proton structure function $F_2(x,Q^2)$ in a wide range 
of $x$ and $Q^2$. 
The NL effects turn out  to be rather weak and amount to 
the $10\%$ correction to $F_2(x,Q^2)$ for $x\lsim 10^{-5}$.  
 Much more pronounced NL effects were found in 
the non-linear  model, described in the literature,  with 
a very soft IR regularization corresponding to 
the IR cutoff at $\simeq \Lambda^{-1}_{QCD}$. The latter issue is also 
commented below.

\end{abstract}

\maketitle


\subsection{Introduction}
The non-perturbative fluctuations of the QCD vacuum
\cite{Shuryak}
 restrict the 
phase space for the 
perturbative (real and virtual) gluons of the BFKL cascade 
\cite{BFKL} thus
 introducing a new scale: 
the correlation/propagation  radius $R_c$ of 
perturbative gluons.  From the fits to 
lattice data 
on field strength correlators $R_c\simeq 0.3$ fm \cite{DIGIACOMO}.
The effects of finite  $R_c$ are consistently  incorporated 
 by the  color dipole (CD) BFKL equation 
\cite{NZZJL94,NZJETP94}.
The perturbative gluons with short propagation length
  do not walk to large distances, 
where they  supposedly  fuse together \cite{BK}. 
The fusion probability appears to be 
controlled by the dimensionless parameter $R_c^2/8B$, where 
$B$ stands for the diffraction cone slope
\cite{FZSLOPE, FZPLB}. 

In this communication we discuss the BFKL \cite{BFKL}, BK \cite{BK}  
phenomenology of DIS
in presence of finite correlation length $R_c$
with particular emphasis on 
a sharpened sensitivity of the non-linear 
effects to the infrared.

\subsection{Color screening of BFKL gluons}
The finite correlation length implies the  Yukawa screened
transverse chromo-electric field of the relativistic quark, 
\beq
{\vec{\cal E}}({\vec\rho})\sim 
g_S(\rho)K_1(\rho/R_c){\vec\rho}/\rho,
\label{eq:EVEC}
\eeq
 where $\rho$ is the
$q-g$ separation. 
The kernel ${\cal K}$ of the CD BFKL equation for the color 
dipole cross section, 
\beq
 {\partial_{\xi} \sigma(\xi,r) } 
={\cal K}\otimes\sigma(\xi,r),\,\,\,\xi=\log(1/x),
\label{eq:BFKL}
\eeq
is related to the flux of the
 Weizs\"acker~-~Williams gluons
$|\vec {\cal E}(\vec\rho)-\vec{\cal E}(\vec\rho+\vec r)|^2$
radiated by the relativistic  $q\bar q$-dipole ${\vec r}$
\cite{NZZJL94,NZJETP94} .
The asymptotic freedom  dictates that $\vec{\cal E}(\vec\rho)$ 
must be  calculated  
with the running QCD charge $g_S(\rho)=\sqrt{4\pi\alpha_S(\rho)}$ 
and 
$\alpha_S(\rho)=4\pi/\beta_0\log(C^2/\rho^2\Lambda^2_{QCD})$, 
where $C=1.5$.

\subsection{DGLAP ordering of dipole sizes and the infrared}
Eq.(\ref{eq:BFKL}),
with  the BK non-linearity \cite{BK} included, 
greatly simplifies for the DGLAP ordering of dipole sizes,
$
r^2\ll\rho^2\ll {  R_c^2},
$
\bea
{\partial_{\xi} \sigma(\xi,r)} =
{C_F\over \pi}\alpha_S(r)r^2
\int_{r^2}^{  R_c^2} {d\rho^2\over \rho^4}\times   
\nonumber\\
\times\left[2\sigma(\xi,\rho)-{\sigma(\xi,\rho)^2\over 8\pi B}\right].
\label{eq:GLRINT}
\eea
Our definition of the profile function in the impact 
parameter $b$-space is
\beq
\Gamma(\xi,r,{\bf b})={\sigma(\xi,r) \over 4\pi { B(\xi,r)}}
\exp\left[-{b^2\over 2{ B}}\right],
\label{eq:GAMMA}
\eeq
where ${B}$ is the diffraction cone slope  and
${d\sigma_{el}/dt}\sim \exp{[{ B}t]}$. Eq.(\ref{eq:GAMMA}) implies that
the unitarity  limit for  $\sigma$ is $8\pi B$. 

The diffraction slope 
 for the forward cone in the dipole-nucleon scattering 
was presented in \cite{SLPL}  in a very symmetric form  
\beq 
B(\xi,r)= {1\over 2}\langle b^2\rangle={1\over 8}r^2+{1\over 3} R_N^2 
+2\alpha^{\prime}_{\Pom}\xi.
\label{eq:BSLOPE}
\eeq
The dynamical 
component of $B$ is given by the last term in Eq. (\ref{eq:BSLOPE}) 
 where 
$\alpha^{\prime}_{\Pom}$ is the Pomeron trajectory slope evaluated first 
in \cite{SL94} (see also \cite{SLPL}). 
Here we only cite the order of magnitude estimate \cite{SLPL} 
\beq
\alpha^{\prime}_{\Pom}\sim
{3 \over 16\pi^{2}} \int d^{2}\vec{r}\,\,\alpha_{S}(r)
R_c^{-2}r^{2}
K_{1}^{2}(r/R_c)  \sim
{3 \over 16\pi}\alpha_{S}(R_{c})R_{c}^{2}
\, ,
\label{eq:ALPRIME}
\endeq
which  clearly shows the connection between the dimensionful
$\alpha^{\prime}_{\Pom}$ and the non-perturbative infrared parameter
$R_{c}$. 
In Eq. (\ref{eq:BSLOPE}) the gluon-probed radius of the proton is a 
phenomenological parameter to be determined from the experiment.
The analysis of Ref. \cite{INS2006} gives
$R_N^2 \approx 12{\, \rm GeV}^{-2}$.

The function 
$ 
\rho^{-2}\sigma(\xi,\rho)\sim\alpha_S(\rho)G(x,\rho),
$
where $G$ is the integrated gluon density, 
is flat in $\rho^2$ and the non-linear term  in Eq.(\ref{eq:GLRINT})
 is dominated by $\rho\sim { R_c}$:
$$
{1\over 8 B}\int_{r^2}^{  R_c^2} {d\rho^2\over \rho^4}
\sigma(\xi,\rho)^2
 \simeq
{  R_c^2\over 8B}\left({\pi^2\over N_c}\alpha_S({  R_c})
G(x,{  R_c})\right)^2.
$$
Thus, the small parameter ${R_c^2/8B}$ enters the game. 
To see it  in action a partial solution to Eq.(\ref{eq:GLRINT})
is needed.

\subsection{Partial solution to GLR-MQ}
The differential form of Eq.(\ref{eq:GLRINT}) for 
$G$ is the GLR-MQ equation \cite{GLRMQ}
\beq
{\partial_{\xi}\partial_{\eta}G(\xi,\eta)} = 
c G(\xi,\eta)-a(\eta)G^2(\xi,\eta),
\label{eq:GLRMQ}
\eeq
where $c=8C_F/\beta_0$, $\eta=\log{[\alpha_S(R_c)/\alpha_S(\rho)]}$,
\beq 
a(\eta)=a(0)\exp[-\eta - \lambda (e^{\eta}-1)],
\label{eq:AETA}
\eeq
$a(0)=\alpha_S(R_c)\pi R_c^{2}/4\beta_0 B$ and
$\lambda=4\pi/\beta_0\alpha_S(R_c)$.
We solve Eq.(\ref{eq:GLRMQ}) making use of the exact solution of
$${\partial_{\xi}\partial_{\eta}G(\xi,\eta)} = 
c G(\xi,\eta)$$
 as a boundary condition. To  simplify
Eq.(\ref{eq:GLRMQ}) we  substitute
 the steeply falling function  $a(\eta)$ with 
$a(\eta)=a(0)\exp{[-\gamma\eta]}$. The partial solution of the 
simplified problem is  found readily:
\beq
G(\xi,\eta)={\exp{(\Delta\xi+\gamma\eta)}\over 
{\omega \exp{(\Delta\xi)}+const}}.
\label{eq:PARTIAL}
\eeq
Here $\gamma=c/\Delta$ with asymptotic value $\Delta=0.4$
 and
\beq
\omega={a(0)\over c}={R_c^2\over 8B}\cdot {\pi\alpha_S(R_c)\over 2N_c}.
\label{eq:OMEGA}
\eeq
One can see that the  gluon fusion mechanism  tames the exponential 
$\xi$-growth of $G(\xi,\eta)$ but fails to stop the accumulation of large 
logarithms, 
$\eta=\log(1/\alpha_S(r))$, at very  small $r^2\ll R_c^2$.

For vanishing non-linearity, $\omega\to 0$, 
Eq.({\ref{eq:PARTIAL}}) matches the exact  solution to the CD  BFKL 
equation found in Ref.\cite{UNIVERSAL}
\beq
G\sim \exp(\Delta\xi+\gamma\eta)\sim 
\left({1\over x}\right)^{\Delta}
\left[1\over \alpha_S(r)\right]^{\gamma}.
\label{eq:WEAKCOUPL}
\eeq
In the limit
$\xi\to \infty  $ the gluon density saturates,
\beq
G\sim \omega^{-1}e^{\gamma\eta},
\label{eq:GUNI}
\eeq
and the corresponding dipole cross section reads
\beq
\sigma(r)=8\pi B\cdot {r^2\over R_c^2}\cdot 2
\left[\alpha_S(R_c)\over\alpha_S(r)\right]^{\gamma}.
\label{eq:SIGMAUNI}
\eeq

\subsection{Regime of the additive quark model} 
At large  $r\gsim R_c$
 a sort of the additive quark model is recovered:
the (anti)quark of the dipole $\vec r$ develops its own perturbative 
gluonic cloud and the pattern of the gluon fusion changes dramatically 
\cite{FZPLB}. From   Ref.\cite{FZPLB} it follows that the  the non-linear 
correction to the dipole cross section is 
\bea
\delta\sigma \sim R_c^{-2}\int^{R_c^2} {d\rho^2}K_1^2(\rho/R_c)
{\sigma(\xi,\rho)\sigma(\xi,r)\over 8\pi B}\sim\nonumber\\
\sim{\sigma(\xi,R_c)\sigma(\xi,r)\over 8\pi B}.
\label{eq:deltaSigma}
\eea
For $R_c$ much smaller than the nucleon size $\sigma(R_c)\propto R_c^2$.
Therefore, the magnitude of 
non-linear effects 
is  controlled, like in the case of small dipoles, by  the ratio 
$R_c^2/8B$.

\subsection{Small $R_c$ - weak non-linearity}
In  Ref.\cite{FZPLB} we solved numerically the BFKL and BK 
equations  with purely perturbative initial conditions and the IR 
regularization described above. Our finding is that 
 the smallness of the ratio $R_c^2/8\pi B$
 makes the non-linear effects rather weak 
even at the lowest  Bjorken $x$ available at HERA.
The linear BFKL with the running coupling and the infrared 
regulator $R_c=0.26$ fm gives a good description of the 
proton structure function $F_2(x,Q^2)$ in a wide range 
of $x$ and $Q^2$ \cite{FZPLB}. For the smallest available 
$x\lsim 10^{-5}$
the $10\%$ NL correction improves the agreement with data, though.

\subsection{Soft IR regularization - fully developed  non-linearity}  
In Ref.\cite{AAA} the non-linear BK-analysis of HERA data on $F_2(x,Q^2)$
was presented. The infrared  cutoff for the purely 
perturbative BK-kernel is deep in the non-perturbative region,
\beq 
r_{IR}=2/Q_s\simeq 4\,{\rm GeV}^{-1}\simeq \Lambda^{-1}_{QCD},
\label{eq:RIR}
\eeq
and the non-linear  effects are  absolutely important 
 to tame the rapid  growth of the linear term in the BK equation.
The running strong coupling $\alpha_S(r)$ at $r=r_{IR}$ in 
 \cite{AAA} appears to be surprisingly small,
$ \alpha_S(r_{IR})\simeq 0.45 $, 
thus assuming  an applicability of
 perturbative QCD to arbitrarily  large distances
 $\sim\Lambda^{-1}_{QCD}$.
However, it is well known that the non-perturbative fields form
 structures with sizes significantly smaller than 
$\Lambda^{-1}_{QCD}$ and local field strength much larger 
than $\Lambda^{2}_{QCD}$. Instantons are one of them \cite{Shuryak}.
Direct confirmation of this picture comes from the lattice 
\cite{DIGIACOMO}. 
Therefore, the approach developed in \cite{AAA}  may lead to a 
very good  BK-description of 
the HERA data  but does not agree with the 
current understanding 
of what is perturbative and non-perturbative in hadronic physics.

\subsection{Summary} 
It is not surprising that introducing a small propagation length for 
perturbative gluons pushes the nonlinear effects to very small $x$.
 More surprising is that within the linear
CD BFKL approach
this small length, $R_c=0.26$ fm,  results in the correct $x$-dependence 
of $F_2(x,Q^2)$ in a wide range of $x$ and $Q^2$. 
The $10\%$ non-linear 
BK-correction improves the agreement with HERA  data at smallest 
available $x$ \cite{FZPLB}.

The non-linear BK-description \cite{AAA} of HERA data
which extends perturbative QCD to  distances
$\simeq \Lambda^{-1}_{QCD}$ contradicts to  well established 
non-perturbative phenomena  and apparently breaks  the hierarchy of scales of soft
 and hard hadronic  physics.

\subsection{Acknowledgments}
V.R.~Z. thanks Alessandro Papa and Agustin Sabio Vera for the kind invitation to
the workshop {\emph Diffraction 2012} and  
support. Thank are due to  Wolfgang Sch\"afer and 
Bronislav Zakharov for useful discussions 
and to  the Dipartimento di Fisica dell'Universit\`a
della Calabria and the INFN - gruppo collegato di Cosenza for their warm
hospitality while a part of this work was done.
The work was supported in part by the Ministero Italiano
dell'Istruzione, dell'Universit\`a e della Ricerca and  by
 the RFBR grants 11-02-00441 and 12-02-00193.





\IfFileExists{\jobname.bbl}{}
 {\typeout{}
  \typeout{******************************************}
  \typeout{** Please run "bibtex \jobname" to optain}
  \typeout{** the bibliography and then re-run LaTeX}
  \typeout{** twice to fix the references!}
  \typeout{******************************************}
  \typeout{}
 }



\end{document}